\documentclass{article}

\usepackage{amsfonts}

\title{Quantum Systems on Linear Groups}

\author{J. J. S\l awianowski\\
Institute of Fundamental Technological Research,\\
Polish Academy of Sciences,\\
21 \'{S}wi\c{e}tokrzyska str., 00-049 Warsaw, Poland\\
e-mail: jslawian@ippt.gov.pl}

\begin{document}

\maketitle
\begin{abstract}

Discussed are quantized dynamical systems on orthogonal and affine
groups. The special stress is laid on geodetic systems with
affinely-invariant kinetic energy operators. The resulting
formulas show that such models may be useful in nuclear and
hadronic dynamics. They differ from traditional Bohr-Mottelson
models where SL$(n,\mathbb{R})$ is used as a so-called
non-invariance group. There is an interesting relationship between
classical and quantized integrable lattices.

\end{abstract}

\noindent {\bf Keywords:} quantized affine systems, quantized
rigid body, multi-valued wave functions.

\section{Introduction. Multi-valued wave functions}

Below we deal with the simple Schr\"odinger quantization, i.e.,
with wave mechanics on differential manifolds.

Let $Q$ be a configuration space, i.e., differential manifold of
dimension $f$ (the number of classical degrees of freedom). If it
is endowed with some positive volume measure $\mu$, then the wave
functions may be considered as complex scalar fields
$\Psi:Q\rightarrow\mathbb{C}$. The corresponding scalar product is
given by
\begin{equation}
\langle\Psi_{1}|\Psi_{2}\rangle=\int\overline{\Psi_{1}(q)}\Psi_{2}(q)d\mu(q),
\end{equation}
and our Hilbert space is meant as $L^{2}(Q,\mu)$. Usually $\mu$
comes from some Riemannian structure $(Q,g)$ and then
$d\mu(q)=\sqrt{|{\rm det}[g_{ij}]|}dq^{1}\cdots dq^{f}$. As shown
and discussed by Mackey \cite{Mackey} one can do quite well
without any $\mu$ if wave amplitudes are considered not as scalars
but instead as complex 1/2-weight densities, and then simply
\begin{equation}
\langle\Psi_{1}|\Psi_{2}\rangle=\int\overline{\Psi_{1}(q)}\Psi_{2}(q)
dq^{1}\cdots dq^{f}.
\end{equation}
The well-known text-book conditions on wave functions (does not
matter sca\-lars or densities) are as follows: (i) $\Psi$ is to be
one-valued all over Q, (ii) $\Psi$ is continuous with derivatives
even at potential jumps. This is justified by: probabilistic
interpretation of $\overline{\Psi}\Psi$, probability conservation,
Sturm-Liouville theory, essential self-adjointness of certain
operators.

There were, however, some arguments by Pauli and Reiss that the
one-valuedness is not the basic postulate of quantum mechanics.
There are some path-dependence phenomena and problems with
globalization of local solutions in multiply-connected $Q$'s.

What is a reasonable "multi-valuedness" in this context? The one
that takes $\overline{Q}$, the universal covering manifold of $Q$
with the projection $\pi:\overline{Q}\rightarrow Q$, and admits
wave functions defined rather on $\overline{Q}$ than on $Q$. This
has also to do with projective representations used in quantum
mechanics. The procedure seems to be reasonable when the co-images
$\pi^{-1}(q)$ are finite sets.

One of possible examples is the system of identical particles,
when removing the diagonals from the Cartesian product and
performing appropriate identifications (taking quotients) one
damages drastically topological structure of the configuration
space.

\section{Rigid body and doubly-valued wave functions}

Another example close to our subject is the rigid body, where
$Q\simeq {\rm SO}(3,\mathbb{R})$ and $\overline{Q}\simeq {\rm
SU}(2)$. Then the projection $\pi$ is $2:1$, i.e., for any $u\in
{\rm SU}(2)$, $\pm u$ are projected onto the same element of
SO$(3,\mathbb{R})$. So, there is a natural hope that the system of
spin-less particles bounded by an appropriate potential making it
(approximately) rigid may show half-integer rotational angular
momentum \cite{ABB95,ABMB95,BBM92,Bar-Racz77,Pau39,Rei39}. By
"spin" in the above "spin-less" we mean the usual intrinsic
angular momentum treated as something  primary, non-explained in
the usual rotational sense. By the way, non-explained need not
mean non-explainable; some idea about fundamental particles as
rigid or deformable quantized tops is often coming back to physics
in spite of its exotic character.

Let $D^{j}:$ SU$(2)\rightarrow$ GL$(2j+1,\mathbb{C})$ denote
irreducible unitary representations of SU$(2)$; $j$ runs over
non-negative integers and half-integers starting from zero (Wigner
matrices). Obviously, $D^{j}(u)^{+}=D^{j}(u)^{-1}=D^{j}(u^{-1})$
and $D^{j}(-u)=(-1)^{2j}D^{j}(u)$. For integer $j$-s $D^{j}$ is
projectable to SO$(3,\mathbb{R})$; for non-integer ones one deals
with "two-valued" representations of SO$(3,\mathbb{R})$.
Traditional symbols for matrix elements are
$D^{j}_{m,m^{\prime}}(u)$, where
$m,m^{\prime}=-j,-j+1,\ldots,j-1,j$. According to the Peter-Weyl
theorem the wave amplitudes on SU$(2)$ may be expanded into
series:
\begin{equation}
\Psi(u)=\sum^{\infty}_{j=0}{\rm
Tr}\left(c^{j}D^{j}\left(u\right)\right),\qquad c^{j}\in {\rm
L}(2j+1,\mathbb{C}).
\end{equation}
Statistical interpretation in SO$(3,\mathbb{R})$,
$\overline{\Psi}\Psi(-u)=\overline{\Psi}\Psi(u)$, imposes the
"superselection" rule: $D^{j}$ with different "halfness" of $j$
cannot be superposed. There are two disjoint situations: the
"fermionic" and "bosonic" ones,
\begin{equation}
\Psi_{f}(u)=\sum^{\infty}_{i=1}{\rm
Tr}\left(c^{i/2}D^{i/2}(u)\right),\qquad
\Psi_{b}(u)=\sum^{\infty}_{j=1}{\rm Tr}\left(c^{j}D^{j}(u)\right).
\end{equation}
Left and right regular translations are described respectively as
\begin{equation}
{}^{\prime}\Psi(u):=\Psi(vu),\quad
{}^{\prime}c^{j}:=c^{j}D^{j}(v),\qquad
{}^{\prime}\Psi(u):=\Psi(uv),\quad
{}^{\prime}c^{j}:=D^{j}(v)c^{j}.
\end{equation}
Let ${\bf L}_{a}$, ${\bf R}_{a}$ be differential operators
generating respectively left and right regular translations,
\begin{eqnarray}
\Psi(u(\overline{\epsilon})u(\overline{k}))&=&\Psi(u(\overline{k}))+\epsilon^{a}
{\bf L}_{a}\Psi(\overline{k})+o(\epsilon),\\
\Psi(u(\overline{k})u(\overline{\epsilon}))&=&\Psi(u(\overline{k}))+\epsilon^{a}
{\bf R}_{a}\Psi(\overline{k})+o(\epsilon),
\end{eqnarray}
where the rotation vector $\overline{k}$ is used, i.e., canonical
coordinates of the first kind,
$u(\overline{k})=\exp\left(-(i/2)k^{a}\sigma_{a}\right)$, $|k|\leq
2\pi$, and $\sigma_{a}$ are, obviously, Pauli matrices.

The laboratory and co-moving representations of spin operators are
given respectively by the following expressions: ${\bf
S}_{a}=(\hbar/i){\bf L}_{a}$, ${\bf \hat{S}}_{a}=(\hbar/i){\bf
R}_{a}$. Their quantum Poisson brackets are as follows:
\begin{equation}
\frac{1}{i\hbar}[{\bf S}_{a},{\bf
S}_{b}]=\varepsilon_{ab}{}^{c}{\bf S}_{c},\qquad
\frac{1}{i\hbar}[{\bf \hat{S}}_{a},{\bf
\hat{S}}_{b}]=-\varepsilon_{ab}{}^{c}{\bf \hat{S}}_{c},\qquad
\frac{1}{i\hbar}[{\bf S}_{a},{\bf \hat{S}}_{b}]=0.
\end{equation}
Obviously,
\begin{equation}\label{a4}
D^{j}(u(\overline{k}))=\exp\left(\frac{i}{2}k^{a}S^{j}_{a}\right),
\end{equation}
where $S^{j}$ are Wigner matrices for the $j$-th angular momentum
\cite{Rose95},
\begin{equation}
\left(S^{j}_{1}\right)^{2}+\left(S^{j}_{2}\right)^{2}+\left(S^{j}_{3}\right)^{2}=\hbar^{2}j(j+1){\rm
Id}_{2j+1}
\end{equation}
(Casimir invariant properties). The action of spin operators on
$D^{j}$ is algebraized: ${\bf S}_{a}D^{j}=S^{j}_{a}D^{j}$,
$c^{j}\mapsto c^{j}S^{j}_{a}$ and ${\bf
\hat{S}}_{a}D^{j}=D^{j}S^{j}_{a}$, $c^{j}\mapsto S^{j}_{a}c^{j}$.
In particular,
\[
{\bf S}_{3}D^{j}_{m,m^{\prime}}=\hbar mD^{j}_{m,m^{\prime}},\qquad
{\bf \hat{S}}_{3}D^{j}_{m,m^{\prime}}=\hbar
m^{\prime}D^{j}_{m,m^{\prime}},
\]
\[
\left(\left({\bf S}_{1}\right)^{2}+\left({\bf
S}_{2}\right)^{2}+\left({\bf
S}_{3}\right)^{2}\right)D^{j}=\left(\left({\bf
\hat{S}}_{1}\right)^{2}+\left({\bf
\hat{S}}_{2}\right)^{2}+\left({\bf
\hat{S}}_{3}\right)^{2}\right)D^{j}=\hbar^{2}j(j+1)D^{j}.
\]
Hamiltonian has the following form:
\begin{equation}
{\bf H}={\bf T}+{\bf V}=\frac{1}{2I_{1}}\left({\bf
\hat{S}}_{1}\right)^{2}+ \frac{1}{2I_{2}}\left({\bf
\hat{S}}_{2}\right)^{2}+\frac{1}{2I_{3}}\left({\bf
\hat{S}}_{3}\right)^{2}+V(u).
\end{equation}
The above ${\bf T}$ is invariant under left regular translations
(spatial rotations). For the spherical top, $I_{1}=I_{2}=I_{3}$,
it is also invariant under right regular translations (material
rotations). For the symmetric top, $I_{1}=I_{2}$, it is invariant
under SO$(2,\mathbb{R})$-right translations (material rotations
about the third main axis of inertia). For the non-degenerate
case, $I_{1}\neq I_{2}\neq I_{3}$, there are no material
symmetries. Nevertheless, for any ratio of inertial moments the
metric tensor underlying the kinetic energy form is left invariant
and so is its induced Riemannian volume. Therefore, this volume is
simply proportional to the Haar measure on SU(2)
(SO$(3,\mathbb{R})$), so it may be directly obtained without
embarrassing manipulations on the anisotropic metric tensor.

It is seen that for the quantized geodetic case, $V=0$, the
problem is fully algebraized and the differential eigenequation
${\bf T}\Psi=E\Psi$ splits into the family of algebraic ones for
$c^{j}$-matrices:
\begin{equation}
\left(\frac{1}{2I_{1}}\left(S^{j}_{1}\right)^{2}+
\frac{1}{2I_{2}}\left(S^{j}_{2}\right)^{2}+
\frac{1}{2I_{3}}\left(S^{j}_{3}\right)^{2}\right)c^{j}=E^{j}c^{j}.
\end{equation}
For the symmetric top, $I_{1}=I_{2}=I$, $I_{3}=K$, and even more
so for the spherical one, $I=K$, $D^{j}_{m,m^{\prime}}$ are
eigenfunctions of the basic operators and of ${\bf T}$ itself, and
the eigenvalues may be immediately found (the degeneracy structure
is explicitly seen):
\begin{equation}
E^{j}_{m^{\prime}}=\frac{\hbar^{2}j(j+1)}{2I}+
\left(\frac{1}{2K}-\frac{1}{2I}\right)\hbar^{2}m^{\prime 2}.
\end{equation}
If $V$ exists and is a simple combination of $D^{j}$-functions,
the problem may be also reduced to algebraic equations on the
basis of Clebsch-Gordan series, however, as a rule the resulting
algebraic system is infinite (thus, in general, effective only
when some approximate truncation is possible).

In the papers \cite{Bert1,Bert2} we considered Bertrand-type
models for the spherical top, i.e., isotropic models with all
trajectories closed. One of them was degenerate oscillator,
$V=2\kappa \tan^{2}(\varphi/2)$, $\kappa>0$, where $\varphi$
denotes the angle of deflection from the equilibrium orientation.
Due to the singularity at $\varphi=\pi$ (infinite potential
barrier) this is, as a matter of fact, the problem on
SO$(3,\mathbb{R})$, the usual rigid body configuration space.
However, in the limit $\kappa\rightarrow 0$, we obtain the free
rigid body with the half-integer angular momentum admitted. Just
another illustration of the idea of "half-integerness" for
extended systems.

Everything said above remains valid for the abstract
$n$-dimensional rigid body, $n>2$, where the $2:1$ covering of
SO$(n,\mathbb{R})$ is the group Spin$(n)$.

\section{Quantized affine bodies and doubly-valued\\ wave functions}

Let us now discuss quantization of an affinely-rigid body
\cite{JJS82,JJS88} without translations. By the affinely-rigid
body we mean such one that all affine relationships between its
constituents are preserved during its motion (rigid in the sense
of affine geometry). So, we deal with Schr\"odinger wave mechanics
on GL$^{+}(n,\mathbb{R})$ or SL$(n,\mathbb{R})$ in the
incompressible case. For $n=3$ such degrees of freedom are used in
the droplet model of atomic nuclei \cite{Bohr-Mot75}. However,
only kinematics is there directly ruled by SL$(3,\mathbb{R})$, the
dynamics is not invariant under this group. Rather,
SL$(3,\mathbb{R})$ is there the dynamical non-invariance group
which enables one to investigate the energy spectrum in terms of
some ladder procedure. The whole beauty and analytical usefulness
of group-theoretic degrees of freedom are then lost. We are going
to construct kinetic energies (metric tensors) invariant under
affine group.

Just as in rigid-body mechanics the most natural Hilbert space
structures are those based on Haar measure $\lambda$ on
GL$^{+}(n,\mathbb{R})$, SL$(n,\mathbb{R})$, i.e., $d\lambda
(\varphi ) =({\rm det} \varphi )^{-n} d l (\varphi ) = ({\rm det}
\varphi )^{-n} \varphi ^{1}{ }_{1} \cdots \varphi ^{n}{ }_{n}$,
where $l$ denoting the usual Lebesgue measure on
L$(n,\mathbb{R})$, i.e., the set of all $n\times n$ real matrices
and, as a matter of fact, the Lie algebra of GL$(n,\mathbb{R})$.
Momentum mappings \cite{Abr-Mars78} corresponding to the left and
right regular translations (laboratory and co-moving
representations) are given by following quantities which may be
meaningfully called affine spin (hypermomentum):
\begin{equation}
{\bf \Sigma}^{a}{}_b =\frac{\hbar}{i} \varphi ^a { }_K
\frac{\partial }{\partial \varphi ^b { }_K } =\frac{\hbar}{i} {\bf
L}^a { }_b,\qquad {\bf \hat{\Sigma}} ^A { }_B =\frac{\hbar}{i}
\varphi ^i { }_B \frac{\partial }{\partial \varphi ^i { }_A }
=\frac{\hbar}{i} {\bf R}^A { }_B.
\end{equation}
They are formally self-adjoint in L$^2({\rm GL}^{+}(n,\mathbb{R}),
\lambda)$ but not on L$^2({\rm GL}^+(n,\mathbb{R}),l)$. To become
such in the latter case they must be completed by some algebraic
terms. Obviously, $\Sigma ^a { }_b = \varphi ^a { }_A {\bf
\hat{\Sigma}} ^A { }_B (\varphi^{-1} )^B { }_b$, and ${\bf L}^a {
}_a $, ${\bf R}^B { }_B$ are generators of the left and right
regular transformations:
\begin{eqnarray}
\Psi((I+\alpha )\varphi )&=& \Psi (\varphi ) + \alpha ^i {}_j
{{\bf L}}^{j}{}_{i} \Psi (\varphi )+ o(\alpha),
\\
\Psi((I+\alpha )\varphi )&=& \Psi (\varphi ) + \alpha ^B {}_A
{{\bf R}} ^{A}{}_{B} \Psi (\varphi )+o(\alpha ).
\end{eqnarray}
The skew-symmetric parts are referred to as spin and vorticity
(Dyson):
\begin{equation}
{\bf S}^i{}_j ={\bf \Sigma}^i{}_i - {\bf \Sigma}_j{}^i ,\qquad
{\bf V}^A{}_B =\hat{\Sigma}^A{}_B - \hat{\Sigma}_B{}^A
\end{equation}
(shift of indices meant in the Kronecker-delta sense). For $n > 2$
the covering group $\overline{{\rm GL} ^+ (n,\mathbb{R})}$ is
$2:1$, and ${\rm GL} ^+ (n,\mathbb{R})$ is doubly-connected.

\noindent{\bf Remark:} $\overline{{\rm GL} ^+ (n,\mathbb{R})}$ is
nonlinear, and so is $\overline{{\rm SL} ^+ (n,\mathbb{R})}$. By
"nonlinear" we mean "non-admitting faithful representations in
terms of finite-dimensional matrices". The doubly-connected
topology of ${\rm GL} ^+ (n,\mathbb{R})$ is seen from the polar
decomposition, ${\rm GL} ^+ (n,\mathbb{R})\ni \varphi =U A$, where
$U\in {\rm SO}(n,\mathbb{R})$ and $A$ is symmetric and positively
definite. For $n\geq 3$, SO$(n,\mathbb{R})$ is doubly-connected,
whereas the manifold of $A$-s has evidently the
$\mathbb{R}^n$-topology. Topologically the covering of ${\rm GL}
^+ (n,\mathbb{R})$ is given by the Cartesian product
Spin$(n,\mathbb{R})\times$ Sym$^+ (n,\mathbb{R})$; in the physical
case $n=3$, just SU(2)$\times$ Sym$^+(3,\mathbb{R})$. And then we
identify skew-symmetric tensors with pseudo-vectors:
\begin{equation}
{\bf S}^i{}_j =\varepsilon^i{}_j{}^k {\bf S}_k,\ \ {\bf
S}_i=\frac{1}{2}\varepsilon _{ij}{}^k {\bf S}^j{}_k,\ \ {\bf
V}^A{}_B =\varepsilon^A{}_B{}^C {\bf V}_C,\ \ {\bf
V}_A=\frac{1}{2}\varepsilon _{AB}{}^C {\bf S}^B{}_C.
\end{equation}
Peter-Weyl expansion on $\overline{{\rm GL} ^+ (3,\mathbb{R})}$
gives us: $\Psi (u,A) = \sum_{s}{\rm Tr}\left(c^s(A) D^s
(u)\right)$, where $s$ are integers and half-integers starting
from $0$. If $\Psi$ is to be admissible as a probabilistically
interpretable wave function on ${\rm GL} ^+ (n,\mathbb{R})$, then
again the "superselection" rule must be satisfied, namely, (i)
only half-integer $s$ are admitted in the series and $\Psi$ is
doubly-valued in ${\rm GL} ^+ (n,\mathbb{R})$, (ii) only integer
$s$ are admitted and $\Psi$ (the more so $\overline{\Psi}\Psi$) is
single-valuated. Moreover, no superposition between (i) and (ii)
is admitted if $\Psi$ is to be statistically interpretable in
GL$^{+}(n,\mathbb{R})$. So, again the "boson-fermion"
superselection rule.

Much more effective, at least in high-symmetry problems, is the
two-polar decomposition GL$^{+}(n,\mathbb{R}) \ni \varphi =
LDR^{-1}$, where $L,R\in$ SO$(n,\mathbb{R})$, and $D$ is diagonal;
it is convenient to write: $D_{aa}=Q^{a}= \exp(q^{a})$. Then
$\varphi$ is represented by $(L,D,R) \in$ SO$(n,\mathbb{R}) \times
\mathbb{R}^{n} \times$ SO$(n,\mathbb{R})$, however, unlike the
polar decomposition, this one is charged with some singularities
and non-uniqueness (although not very embarrassing when carefully
treated). The Haar and Lebesgue measure $\lambda$, $l$ are then
represented as follows \cite{Bar-Racz77}:
\begin{eqnarray}
d \lambda (\varphi) &=& d \lambda \left(L,q,R\right)=
P_{\lambda}(q) d\mu(L) d\mu(R) dq^{1} \cdots dq^{n},\\ d l
(\varphi) &=& d l \left(L,Q,R\right)= P_{l}(Q) d\mu(L) d\mu(R)
dQ^{1} \cdots dQ^{n},
\end{eqnarray}
where $\mu$ is the Haar measure on SO$(n,\mathbb{R})$ and
\begin{equation}
P_{\lambda}(q)= \prod_{i \neq j} \left| \sinh
\left(q^{i}-q^{j}\right)\right|, \quad P_{l}(Q)= \prod_{i \neq j}
\left|\left(Q^{i}-Q^{j}\right)\left(Q^{i}+Q^{j}\right)\right|.
\end{equation}
${\mathbf{S}^{i}}_{j}$, ${\mathbf{V}^{A}}_{B}$ generate left
SO$(n,\mathbb{R})$-regular translations of $L,R$-factors. Right
regular translations are generated respectively by
${\mathbf{\rho}^{a}}_{b}=
\left(L^{-1}\right)^{a}{}_{i}{\mathbf{S}^{i}}_{j}{L^{j}}_{b}$ and
${\mathbf{\tau}^{a}}_{b}=\left(R^{-1}\right)^{a}{}_{A}
{\mathbf{V}^{A}}_{B}{R^{B}}_{b}$. As usual, for $n=3$ we represent
them as follows:
\begin{equation}
{\mathbf{\rho}^{a}}_{b}=\epsilon^{a}{}_{b}{}^c
{\mathbf{\rho}}_{c}, \quad
{\mathbf{\tau}^{a}}_{b}=\epsilon^{a}{}_{b}{}^c
{\mathbf{\tau}}_{c}, \quad
{\mathbf{\rho}}_{a}=\frac{1}{2}\epsilon_{ab}{}^c
{\mathbf{\rho}}^b{}_{c},\quad
{\mathbf{\tau}_{a}}=\frac{1}{2}\epsilon_{ab}{}^c
{\mathbf{\tau}}^b{}_{c}.
\end{equation}
To deal with the doubly-valued functions, i.e., with the covering
manifold, we begin with SU$(2)\times \mathbb{R}^3\times$ SU$(2)$
as an auxiliary tool. The Peter-Weyl theorem gives us the
following expansion:
\begin{equation}
\Psi (u,q,v)=\sum_{s,j} \sum ^{s} _{m,n=-s}    \sum^{j} _{k,l=-j}
D^s{}_{mn}  (u)  f^{sj}_{{}^{nk}_{ml}}(q) D^j{}_{kl}
\left(v^{-1}\right),
\end{equation}
or for eigenstates of ${\bf S}_3$, ${\bf V}_3$ with $\hbar m,\hbar
l$-eigenvalues:
\begin{equation}
\Psi ^{sj} _{ml} (u,q,v)=\sum^{s}_{n=-s} \sum ^{j} _{k=-j}
 D^s{}_{mn}  (u)  f^{sj}_{nk}  (q)
D^j{}_{kl} \left(v^{-1}\right).
\end{equation}
Obviously, here $\hbar^2 s(s+1)$ and $\hbar^2 j(j+1)$ are
eigenvalues of ${\bf S}$- and ${\bf V}$-Casimirs. But SU$(2)\times
\mathbb{R}^3\times{\rm SU}(2)$ is not diffeomorphic with
$\overline{{\rm GL}^{+}(3,\mathbb{R})}$. One can show that the
above expressions are well-defined wave functions on
$\overline{{\rm GL}^{+}(3,\mathbb{R})}$, i.e., "good"
doubly-valued wave functions on GL$(3,\mathbb{R})$ if $(j-s)$ is
an integer, i.e., $j$ and $s$ have the same "halfness". Besides,
some additional conditions must be satisfied to take into account
that the two-polar decomposition of GL$^{+}(3,\mathbb{R})$ in
terms of SO$(3,\mathbb{R})\times\mathbb{R}^3\times{\rm
SO}(3,\mathbb{R})$ is non-unique \cite{JJS82,JJS88}. The above
wave functions are single-valued in GL$(3,\mathbb{R})$, when $s$
and $j$ are integers. So, $\sum_{s,j:(j-s)\in \mathbb{Z}}$ and
$\sum_{s,j \in \mathbb{N}\cup \{0\}}$ are well-defined
respectively on $\overline{{\rm GL}(n,\mathbb{R})}$ and
GL$^{+}(n,\mathbb{R})$. And again there is the "superselection":
the latter sum can not be combined with $\sum_{s,j \in
\mathbb{N}/2}$.

We are interested in affinely-invariant geodetic models, i.e., in
free affine top. Let us stress, however, that strictly speaking
purely geodetic model would be non-physical because it would
predict the non-limited dilatational expansion and collapse
(although in the infinite time). Therefore, the logarithmic
dilatational parameter $q =
\left(q^{1}+q^{2}+\cdots+q^{n}\right)/n$ ($n=3$ in the physical
case; sometimes $n=2$ or $n=1$) must be stabilized by some simple
model potential $V(q)$, e.g., harmonic oscillator $V_{\rm
osc}=(\kappa/2)q^{2}$ or the infinite potential well. It turns out
that the incompressible (thus applicable in nuclear and hadronic
dynamics) geodetic SL$(n,\mathbb{R})$-models are realistic both on
the classical and quantum level and may predict bounded vibrating
behaviour (and below-threshold discrete spectrum in quantum
theory). In analogy to the spherical rigid body we can postulate
the left and right invariant kinetic energy on
GL$^{+}(n,\mathbb{R})$. On the classical level it would be given
by the Casimir expression:
\begin{equation}
T = \frac{A}{2} {\rm Tr} \left( \Omega^{2} \right)+ \frac{B}{2}
\left( {\rm Tr} \Omega \right)^{2}=\frac{A}{2} {\rm Tr} \left(
\hat{\Omega}^{2} \right)+ \frac{B}{2} \left({\rm Tr} \hat{\Omega}
\right)^{2},
\end{equation}
where $\Omega$, $\hat{\Omega}$ are Lie-algebraic objects, just the
affine counterparts of the laboratory and co-moving
representations of the angular velocity,
\begin{equation}
\Omega = \frac{d\varphi}{dt} \varphi^{-1}, \qquad \hat{\Omega} =
\varphi^{-1} \frac{d\varphi}{dt} = \varphi^{-1} \Omega \varphi.
\end{equation}
The corresponding Laplace-Beltrami operator is expressed in the
two-polar terms as follows (in $n$ dimensions):
\begin{eqnarray}
\mathbf{T}^{\rm aff-aff}=&-&\frac{\hbar^{2}}{2A}
\mathbf{D}_{\lambda} +
\frac{\hbar^{2}B}{2A(A+nB)}\frac{\partial^{2}}{\partial q^{2}}
\nonumber\\
&+&\frac{1}{32A} \sum_{a,b}\frac{\left({\bf \rho}^{a}{}_{b} -
{{\mathbf{\tau}}^{a}}_{b}\right)^{2}}{\sinh^{2}\frac{q^{a}-q^{b}}{2}}-\frac{1}{32A}
\sum_{a,b}\frac{\left({{\mathbf{\rho}}^{a}}_{b}
+{{\mathbf{\tau}}^{a}}_{b}\right)^{2}}{\cosh^{2}\frac{q^{a}-q^{b}}{2}}.\nonumber
\end{eqnarray}
The differential operator $\mathbf{D}_{\lambda}$ is given by the
following expression:
\begin{equation}
\mathbf{D}_{\lambda}= \frac{1}{P_{\lambda}} \sum_{a}
\frac{\mathbf{\partial}}{\mathbf{\partial}q^{a}}
P_{\lambda}\frac{\mathbf{\partial}}{\mathbf{\partial}q^{a}}.
\end{equation}
This kinetic energy is not positively definite, but its negative
term may encode the attraction (strange "centrifugal" attraction)
of deformation invariants, whereas its positive counterpart
describes the repulsive forces (infinite at coincidence
situation). Their balance leads on the classical level to
nonlinear elastic vibrations with an open subset of bounded
trajectories and an open subset of non-bounded ("dissociated")
ones; everything, of course, under the assumption of approximate
incompressibility, when some dilatation-stabilizing potential
$V(q)$ is used. On the quantum level this means that both the
discrete spectrum and the higher-placed continuous one do occur.

For certain reasons it may be convenient to  discuss models
$\mathbf{T}^{\rm met-aff}$ invariant under spatial rotations (left
translations by orthogonal elements) and right affine
transformations, and also conversely, the models $\mathbf{T}^{\rm
aff-met}$ with opposite symmetry properties. Classically:
\begin{eqnarray}
T^{\rm met-aff} &=& \frac{I}{2} {\rm Tr} \left(
\Omega^{T}\Omega\right)+ \frac{A}{2} {\rm Tr} \left(
\Omega^{2}\right)+\frac{B}{2} \left( {\rm Tr} \Omega\right)^{2},
\\
T^{\rm aff-met}&=& \frac{I}{2} {\rm Tr} \left( \hat{\Omega}^{T}
\hat{\Omega}\right)+ \frac{A}{2} {\rm Tr} \left(
\hat{\Omega}^{2}\right)+\frac{B}{2} \left( {\rm Tr}
\hat{\Omega}\right)^{2}.
\end{eqnarray}
The first one is a discretization of the Arnold model of ideal
fluid as a Hamiltonian system on the group of volume-preserving
diffeomorphisms. The second one does not obey the spatial metric
relations like, e.g., electrons in crystals, for which the metric
tensor is replaced by the effective mass tensor; similar things
happen in the theory of defects in solids. After quantization:
\begin{eqnarray}
\mathbf{T}^{\rm met-aff} &=& \mathbf{T}^{\rm aff-aff}\left[A
\mapsto I+A\right]+ \frac{I}{2\left(I^{2}-A^{2}\right)}
\|\mathbf{S}\|^{2},
\\
\mathbf{T}^{\rm aff-met} &=& \mathbf{T}^{\rm aff-aff}\left[A
\mapsto I+A\right]+ \frac{I}{2\left(I^{2}-A^{2}\right)}
\|\mathbf{V}\|^{2}.
\end{eqnarray}
The shorthand $A \mapsto I+A$ means obviously "with $A$ replaced
by $I+A$"; $\|\mathbf{S}\|^{2}$ and $\|\mathbf{V}\|^{2}$ are
squared magnitudes of spin and vorticity, i.e., Casimirs:
\begin{equation}
\|\mathbf{S}\|^{2}=-\frac{1}{2}
{\mathbf{S}^{a}}_{b}{\mathbf{S}^{b}}_{a}, \quad
\|\mathbf{V}\|^{2}=-\frac{1}{2}
{\mathbf{V}^{A}}_{B}{\mathbf{V}^{B}}_{A}.
\end{equation}
For the proper choice of $I$, $A$, $B$, these  kinetic energies
are positively definite. For the dynamically non-affine but
physically-justified macroscopically-elastic models with double
isotropy, $T^{\rm d'A}=(I/2){\rm Tr} \left(\dot{\varphi}^{T}
\dot{\varphi}\right)$, we have
\begin{equation}
\mathbf{T}^{\rm d'A}= -\frac{\hbar^{2}}{2I} \mathbf{D}_{l}+
\frac{1}{8I} \sum_{a,b}\frac{\left({{\mathbf{\rho}}^{a}}_{b}
-{{\mathbf{\tau}}^{a}}_{b}\right)^{2}}{\left(Q^{a}-Q^{b}\right)^{2}}+
\frac{1}{8I} \sum_{a,b}\frac{\left({{\mathbf{\rho}}^{a}}_{b}
+{{\mathbf{\tau}}^{a}}_{b}\right)^{2}}{\left(Q^{a}+Q^{b}\right)^{2}},
\end{equation}
where
\begin{equation}
{\bf D}_{l}=\frac{1}{P_{l}}\sum_{a}\frac{\partial}{\partial
q^{a}}P_{l}\frac{\partial}{\partial q^{a}}.
\end{equation}
Without potential, the above geodetic model is non-physical. There
are only purely decaying, scattering motions. With some
well-adapted potentials invariant under left and right orthogonal
translations such a model is useful in macroscopic and molecular
problems without, however, any advantage typical for invariant
geodetic systems on groups.

For geodetic models and, more generally, for the doubly-isotropic
potential models, $V=V(q^{1}, \ldots, q^{n})$ (including those
SL$(n,\mathbb{R})$-geodetic with $V(q)$ stabilizing dilatations),
the quantities ${\bf S}^{2}={\bf \varrho}^{2}$, ${\bf V}^{2}={\bf
\tau}^{2}$ are constants of motion, and $s$, $j$ are good quantum
numbers. Then, for fixed $s$, $j$ the stationary Schr\"odinger
equation splits into the family of reduced Schr\"odinger equations
for matrix amplitudes $f^{sj}$ depending only on deformation
invariants $q^{1}, \ldots, q^{n}$ (by deformation invariants one
means, in general, the functions of the matrix $\varphi$ invariant
under left and right regular translations by the orthogonal group
SO$(n,\mathbb{R})$); the dependence on angles is algebraized:
${\bf H}^{sj}f^{sj}=E^{sj}f^{sj}$. For the affine-affine model the
reduced Hamiltonian has the form
\begin{eqnarray}
{\bf H}_{\rm aff-aff}^{sj}f^{sj}=&-&\frac{\hbar ^{2}}{2A}{\bf
D}f^{sj}+\frac{\hbar ^{2}B}{2A(A+nB)}\frac{\partial
^{2}f^{sj}}{\partial q^{2}}+V\left(q^{a}\right)f^{sj}\nonumber\\
&+&\frac{1}{32A}\sum_{a,
b}\frac{\left(\overleftarrow{S^{j}_{ab}}-\overrightarrow{S^{s}_{ab}}\right)^{2}}
{\sinh^{2}\frac{q^{a}-q^{b}}{2}}f^{sj}-\frac{1}{32A}\sum_{a,
b}\frac{\left(\overleftarrow{S^{j}_{ab}}+\overrightarrow{S^{s}_{ab}}\right)^{2}}
{\cosh^{2}\frac{q^{a}-q^{b}}{2}}f^{sj},\nonumber
\end{eqnarray}
where $\overleftarrow{S^{j}_{ab}}f^{sj}:=f^{sj}S^{j}_{ab}$,
$\overrightarrow{S^{s}_{ab}}f^{sj}:=S^{s}_{ab}f^{sj}$. The symbols
$s$, $j$ suggest the dimension $n=3$ and the usual range of these
quantum numbers. Nevertheless, the formula may be meant for the
general $n$, then $s$, $j$ simply run over the set of labels of
irreducible unitary representations of SO$(n, \mathbb{R})$, and
$S^{s}_{ab}$, $S^{j}_{ab}$ are basic (hermitian) generators of
these representations (\ref{a4}). For $n=3$ they are simply the
standard Wigner matrices of angular momentum. The potential $V$ is
necessary only for stabilization or constraining dilatations; on
SL$(n, \mathbb{R})$ the potential-free geodetic model is
satisfactory.

For the metric-affine model the reduced Hamiltonian is given by
\begin{equation}
{\bf H}_{\rm met-aff}^{sj}={\bf H}_{\rm aff-aff}^{sj}[A \mapsto
I+A] +\frac{I}{2(I^{2}-A^{2})}\hbar^{2}C(2,s),
\end{equation}
where $-C(2,s)$ is the eigenvalue of the second-order Casimir
invariant built of generators of regular translations on SO$(n,
\mathbb{R})$ in the $s$-th representation:
\begin{equation}
\frac{1}{2}{\bf L}^{a}{}_{b}{\bf L}^{b}{}_{a}D^{s}=C(2, s)D^{s},
\quad {\rm i.e.,}\quad \frac{1}{2}\sum_{a,
b}S^{s}_{ab}S^{s}_{ba}=\hbar^{2}C(2, s)I_{N},
\end{equation}
where $I_{N}$ denotes the $N \times N$ unit matrix, and $N$ is the
dimension of the $s$-th irreducible representation of SO$(n,
\mathbb{R})$. Obviously, in the interesting physical case $n=3$,
$N=2s+1$, $s=0,1/2,1,\ldots$, and $C(2, s)=s(s+1)$.

For the affine-metric model we have
\begin{equation}
{\bf H}_{\rm aff-met}^{sj}={\bf H}_{\rm aff-aff}^{sj}[A \mapsto
I+A] +\frac{I}{2(I^{2}-A^{2})}\hbar^{2}C(2, j).
\end{equation}
In a sense, ${\bf H}_{\rm aff-aff}^{sj}$ and occurrence of
additional terms proportional to $\hbar^{2}C(2, s)$,
$\hbar^{2}C(2, j)$ is extremely interesting and seems to be
confirmed by the nuclear and hadronic experimental data
respectively as the angular momentum and isospin terms. In the
incompressible  case, when $B=0$, the quantized geodetic model
(without potential) is sufficient for predicting both the discrete
and continuous spectrum (bounded and decaying situations). In the
compressible case, dilatations must be stabilized by some model
potential $V(q)$. Appearing of the discrete and continuous spectra
is controlled by the interplay between $s$ and $j$ quantum numbers
(they are good quantum numbers corresponding to quantum constants
of motion).

The appearance of the formal similarity of the above expressions
to integrable lattices formulas is not accidental and may be
helpful in the analysis of Sutherland and Calogero-Moser lattices.

\section{Some final remarks}

Linear group GL$(3,\mathbb{R})$ has been used in nuclear physics
as the group which rules geometry of the collective degrees of
freedom in the droplet model of nuclei \cite{Bohr-Mot75,Ros-Tr98}.
However, it was not there the group of dynamical symmetries
preserving the Hamiltonian. There are models where
GL$(3,\mathbb{R})$ is the so-called non-invariance group. We
suggest models which seem to be viable and use GL$(3,\mathbb{R})$
as the group of dynamical symmetries.

\subsection*{Acknowledgements}

The author is greatly indebted to the Organizers of the IQSA
Conference, Denver 2004, first of all to professor Franklin
Schroeck Junior, for their cordial hospitality and financial
support.

\end{document}